\documentclass{article}
\usepackage{amsmath}
\usepackage{amssymb}
\usepackage{graphicx}
\title{
\textbf{
Gauging gravity with $SO(1,3)$ for spin-$1/2$ particles}}
\author{Arpan Saha$^\dag$\\ Banibrata Mukhopadhyay$^\ddag$}
\begin{document}
\maketitle
\begin{abstract}
We demonstrate, by analogy with electromagnetism, that the geometric content in the theory of gravity is an indirect consequence of the fact that the gauge group in question is the Lorentz group $SO(1,3)$. We hence 
construct field equations for gravity and a spin-1/2 particle in a gravitational field based on gauge considerations. Furthermore, we derive the weak field and Schr\"odinger limits of the Dirac equation of the particle in the gravitational field, especially in Fermi normal coordinates and on the equatorial plane of the
Kerr geometry, following which we identify the terms to which the electromagnetic potentials $\vec{A}$ and $\Phi$ are analogous. 
\end{abstract}
\let\thefootnote\relax\footnotetext{\begin{flushleft}
$\dag$ Department of Physics, Indian Institute of Technology Bombay, Mumbai; arpan$\_$saha@iitb.ac.in\\
$\ddag$ Department of Physics, Indian Institute of Science, Bangalore; bm@physics.iisc.ernet.in
\end{flushleft}}
\section{Introduction}
One of the principal attractions of Einstein's theory of gravity is the fact that it is geometric in nature -- the gravitational field, instead of being regarded as a force field acting upon particles, is seen as being manifested as the intrinsic curvature of the spacetime semi-Riemannian manifold. As geometry forms one of the fundamental arenas wherein we historically learnt how to exercise logic, it was hoped that such geometric insight would carry over to theories describing the other three forces of Nature -- electromagnetic, weak and strong. This unfortunately turned out not to be the case. As a result, many physicists \cite{Weinberg} started 
to regard treatments of gravity that are geometric in nature from the outset with a certain level of suspicion. Building the theory from the principle of equivalence proves to be a better strategy as it is a far more reliable connect amongst the four forces than geometry.
\\

In the above spirit, this paper takes up the semi-classical description of a spin-1/2 particle in a gravitational field and shows that much of the information about the dynamics of particles and fields can be obtained from gauge considerations alone. By this we mean that if we require that the dynamics on spacetime remains unchanged when the connection associated with the principal bundle corresponding to the gauge group in question undergoes a gauge transformation, we can effectively reconstruct the governing dynamical equations or at least impose constraints strong enough on their form, so that straightforward arguments may be used to eliminate the candidates which are clearly not physical. The group associated with the internal space of symmetry in the case of electromagnetism is $U(1)$, while that in the case of gravity is the Lorentz group $SO(1,3)$. This is what imparts geometric content to the theory of gravity i.e.~the fact that information about the gravitational field is encoded in the metric.
\\

Thus, our twofold claim is as follows.
\begin{enumerate}
\item The fact that the metric carries information about the gravitational field is an indirect consequence of the fact that the gauge group is $SO(1,3)$ and certain other conditions of physicality, as we shall see.
\item The field equations for gravity may be constructed through gauge considerations without any \emph{a priori} geometric assumptions.
\end{enumerate}
Following the demonstration of the above, we shall take up the Dirac equation for a particle in a gravitational field that we will obtain as a result of our analysis, and investigate its Klein-Gordon form and the limits thereof in the weak field and Schr\"odinger regime (when the velocity of the particle is much lower than the speed of light). We finally conclude by looking at the magnetic analogue of the gravitational potential 
in Fermi normal coordinates and on the equatorial plane of the Kerr geometry.

\section{Electromagnetism and $U(1)$}

In this section, we will recall how the dynamical equations for an electromagnetic field and a spin-1/2 particle in it may be developed, given that the gauge group in question is $U(1)$. The motive is to illustrate the key steps in our derivation (which will carry over to the case for gravity) in a setting that we understand relatively well and which is not geometric in nature. This will enable us to better highlight the analogies and differences between electromagnetism and gravity. 
\\

We enumerate the aforementioned steps as follows.

\begin{enumerate}
\item The Lagrangian density for the particle and the field is written as a sum of two parts -- the Lagrangian density for the particle and the Lagrangian density for the field. While the former may be obtained by simply replacing the derivative in the no-field Lagrangian density by a `gauge covariant derivative' wherein a connection term is included so that we may compare objects lying in fibers over two different points, the latter is not known a priori.

\item The elements of the gauge group are allowed to act on the wavefunction spinors (and other objects dwelling in the internal space of symmetry). The particle Lagrangian density must remain invariant under such transformations; this gives us the transformation of the connection.

\item The field Lagrangian density is stipulated to be a gauge-invariant function of the connection. Scalars constructed out of the `gauge curvature' (i.e.~commutators of the gauge covariant derivatives) fit this requirement rather well.

\item The resulting action is varied around the stationary `points'. This yields the equations governing the behaviour of the particle and the field.
\end{enumerate}

First the Lagrangian density $\mathcal{L}$ is split as
\begin{equation}
\mathcal{L} = \mathcal{L}_{1/2} + \mathcal{L}_{EM}
\end{equation}
where $\mathcal{L}_{EM}$ is the Lagrangian density of the field, whose form is to be determined, and $\mathcal{L}_{1/2}$ is the Lagrangian density of the particle given by \cite{Itzykson}
\begin{equation}
\mathcal{L}_{1/2}= \frac{i}{2} \left[\bar\psi\gamma^\mu(\psi_{,\mu} - iqA_\mu\psi) - (\bar\psi_{,\mu}+iq\bar\psi A_\mu)\gamma^\mu\psi \right] - m \bar\psi \psi
\end{equation}
where $\psi$ is the wavefunction spinor, $\bar{\psi}$ is its adjoint spinor, $m$ and $q$ are the mass and charge of the particle respectively, $\gamma^\mu$ are the metric-dependent Dirac matrices and $A_\mu$ is the electromagnetic connection field i.e.~the $4$-vector potential.
\\

Now, the elements of the gauge group $U(1)$ are represented by $e^{iq\chi}$. Therefore, under the action of such an element, the following transformations occur: $\psi\rightarrow\psi^\prime$, $\bar{\psi}\rightarrow\bar{\psi}^\prime$ and $A_\mu\rightarrow A_\mu^\prime$, where
\begin{subequations}
\begin{align}
\psi^\prime &= e^{iq\chi}\psi\\
\bar{\psi}^\prime &= \bar{\psi} e^{-iq\chi}
\end{align}
\end{subequations}
Moreover, the Dirac matrices $\gamma^\mu$ act as mapping operators from physical spacetime (i.e.~the tangent bundle $T\mathcal{M}$ of the semi-Riemannian manifold $\mathcal{M}$ that is spacetime) to the internal space of symmetry associated with each point, and hence must transform as well when the internal space of symmetry is acted upon by the group. Therefore, we have
\begin{equation}\tag{3c}
\gamma^{\prime\mu} = e^{iq\chi}\gamma^\mu e^{-iq\chi}
\end{equation}
Of course, here since the Dirac matrices commute with the elements of $U(1)$, this does not change anything. But this shall not be the case with gravity, as we shall see in the next section.
\\

The result of the above transformations is
\begin{align}
\mathcal{L}_{1/2}^\prime &= \frac{i}{2} \left[\bar\psi^\prime\gamma^{\prime\mu}(\psi^\prime_{,\mu} - iqA^\prime_\mu\psi^\prime) - (\bar\psi^\prime_{,\mu}+iq\bar\psi^\prime A_\mu^\prime)\gamma^{\prime\mu}\psi^\prime \right] - m \bar\psi^\prime \psi^\prime\notag\\
&= \frac{i}{2} \left[\bar{\psi} e^{-iq\chi}e^{iq\chi}\gamma^\mu e^{-iq\chi}((e^{iq\chi}\psi)_{,\mu} - iqA^\prime_\mu e^{iq\chi}\psi)\right.\notag\\
  &\qquad -\left.((\bar{\psi} e^{-iq\chi})_{,\mu}+iq\bar{\psi} e^{-iq\chi} A_\mu^\prime)e^{iq\chi}\gamma^\mu e^{-iq\chi}e^{iq\chi}\psi \right] - m \bar{\psi} e^{-iq\chi} e^{iq\chi}\psi\notag\\
&= \frac{i}{2} \left[\bar{\psi} \gamma^\mu(iq\chi_{,\mu}\psi+\psi_{,\mu} - iqA^\prime_\mu \psi) -(\bar{\psi}_{,\mu} - iq\bar\psi \chi_{,\mu}+iq\bar{\psi}  A_\mu^\prime)\gamma^\mu \psi \right] - m \bar{\psi}\psi
\end{align}
Requiring the Lagrangian density $\mathcal{L}_{1/2}$ to be invariant under such transformations (i.e.~$\mathcal{L}_{1/2}=\mathcal{L}_{1/2}^\prime$) yields the following transformation law for $A_\mu$
\begin{equation}
A_\mu^\prime = A_\mu + \chi_{,\mu}
\end{equation}
The gauge curvature $F_{\mu\nu} = A_{\nu,\mu} - A_{\mu,\nu}$ effectively takes care of the $\chi_{,\mu}$ term, and we stipulate that the field Lagrangian density is a scalar constructed out of $F_{\mu\nu}$. We also stipulate, out of empirical considerations, that the constructed scalar minimises the number of contractions taking place while not identically vanishing (as otherwise, that would lead to loss of all physical information). 
In the present case, this is quite straightforward and given by
\begin{equation}
\mathcal{L}_{EM} = K g^{\mu\lambda}g^{\nu\sigma}F_{\mu\nu}F_{\lambda\sigma}
\end{equation}
where $g^{\mu\lambda}$ denotes the metric tensor and $K$ is a constant of proportionality.
\\

The action, hence, is
\begin{align}
S &= \int \mathcal{L} \sqrt{-g}~d^4x\notag\\
&= \int \left(\mathcal{L}_{1/2}+\mathcal{L}_{EM}\right)\sqrt{-g}~d^4x\notag\\
&= \int \left(\frac{i}{2} \left[\bar\psi\gamma^\mu(\psi_{,\mu} - iqA_\mu\psi) - (\bar\psi_{,\mu}+iq\bar\psi A_\mu)\gamma^\mu\psi \right]\right.\notag\\
&\qquad \left.\indent\indent\indent-~m \bar\psi \psi+K g^{\mu\lambda}g^{\nu\sigma}F_{\mu\nu}F_{\lambda\sigma}\right)\sqrt{-g}~d^4x
\end{align}
Here, $g = \det g^{\mu\nu}$ and the factor $\sqrt{-g}$ is included to account for the fact that the infinitesimal volume $d^4x$ does not transform like a scalar under general coordinate transformations. 
\\

Here, the independent variables are taken to be $\psi$, $\bar{\psi}$ and $A_\mu$. The Euler-Lagrange 
equations are then obtained as
\begin{subequations}
\begin{align}
\partial_\mu\left(\frac{\partial\left(\mathcal{L}\sqrt{-g}\right)}{\partial\psi_{,\mu}}\right) - \frac{\partial\left(\mathcal{L}\sqrt{-g}\right)}{\partial\psi} &= 0\\
\partial_\mu\left(\frac{\partial\left(\mathcal{L}\sqrt{-g}\right)}{\partial\bar\psi_{,\mu}}\right) - \frac{\partial\left(\mathcal{L}\sqrt{-g}\right)}{\partial\bar\psi} &= 0\\
\partial_\mu\left(\frac{\partial\left(\mathcal{L}\sqrt{-g}\right)}{\partial A_{\nu,\mu}}\right) - \frac{\partial\left(\mathcal{L}\sqrt{-g}\right)}{\partial A_\nu} &= 0
\end{align}
\end{subequations}
We may use the fact that $\left(\partial_\mu\sqrt{-g}\right)=\frac{1}{2}\sqrt{-g}g^{\lambda\nu}g_{\lambda\nu,\mu}$ to factor out $\sqrt{-g}$ and obtain
\begin{subequations}\label{ELemag}
\begin{align}
\frac{1}{2}g^{\lambda\rho}g_{\lambda\rho,\mu}\left(\frac{\partial\mathcal{L}}{\partial\psi_{,\mu}}\right)+\partial_\mu\left(\frac{\partial\mathcal{L}}{\partial\psi_{,\mu}}\right) - \frac{\partial\mathcal{L}}{\partial\psi} &= 0\\
\frac{1}{2}g^{\lambda\rho}g_{\lambda\rho,\mu}\left(\frac{\partial\mathcal{L}}{\partial\bar\psi_{,\mu}}\right)+\partial_\mu\left(\frac{\partial\mathcal{L}}{\partial\bar\psi_{,\mu}}\right) - \frac{\partial\mathcal{L}}{\partial\bar\psi} &= 0\\
\frac{1}{2}g^{\lambda\rho}g_{\lambda\rho,\mu}\left(\frac{\partial\mathcal{L}}{\partial A_{\nu,\mu}}\right)+\partial_\mu\left(\frac{\partial\mathcal{L}}{\partial A_{\nu,\mu}}\right) - \frac{\partial\mathcal{L}}{\partial A_\nu} &= 0
\end{align}
\end{subequations}
We now compute the derivatives as
\begin{align*}
\frac{\partial\mathcal{L}}{\partial\psi} &= \frac{i}{2}\left[ -iq\bar\psi\gamma^\mu A_\mu - (\bar\psi_{,\mu}+iq\bar\psi A_\mu)\gamma^\mu\right]  - m \bar\psi\\
\frac{\partial\mathcal{L}}{\partial\bar\psi} &= \frac{i}{2} \left[\gamma^\mu(\psi_{,\mu} - iqA_\mu\psi) - iq A_\mu\gamma^\mu\psi \right] - m\psi\\
\frac{\partial\mathcal{L}}{\partial A_\nu}&=q\bar\psi\gamma^\nu\psi\\
\frac{\partial\mathcal{L}}{\partial\psi_{,\mu}}&=\frac{i}{2}\bar\psi\gamma^\mu \\
\frac{\partial\mathcal{L}}{\partial\bar\psi_{,\mu}}&=-\frac{i}{2}\gamma^\mu\psi \\
\frac{\partial\mathcal{L}}{A_{\nu, \mu}} &= Kg^{\alpha\lambda}g^{\beta\rho}\left[\left(\delta^\nu_\beta\delta^\mu_\alpha-\delta^\nu_\alpha\delta^\mu_\beta\right)F_{\lambda\rho}+F_{\alpha\beta}\left(\delta^\nu_\rho\delta^\mu_\lambda-\delta^\nu_\lambda\delta^\mu_\rho\right)\right]\\
&= 4KF^{\mu\nu}
\end{align*}
The Euler-Lagrange equations (9) thus boil down to the following three equations
\begin{subequations}
\begin{align}
i (\bar\psi_{,\mu}+ iq\bar\psi A_\mu)\gamma^\mu + m\bar\psi &= -\frac{i}{4}g^{\lambda\rho}g_{\lambda\rho,\mu}\bar\psi\gamma^\mu\\
i\gamma^\mu (\psi_{,\mu} - iqA_\mu\psi) - m\psi &= -\frac{i}{4}g^{\lambda\rho}g_{\lambda\rho,\mu}\gamma^\mu\psi\\
F^{\mu\nu}_{~~,\mu}-\frac{q}{4K}\bar\psi\gamma^\nu\psi&=-\frac{1}{2}g^{\lambda\rho}g_{\lambda\rho,\mu}F^{\mu\nu}
\end{align}
\end{subequations}
The three equations we have obtained are in fact the adjoint Dirac equation, the Dirac equation and the source-dependent Maxwell equation (the source-free one becomes tautologous) respectively. The terms on the right side of the equality represent corrections accounting for (possibly) noncartesian coordinates. All this is consistent with what we already know about the dynamics of a spin-1/2 particle in an electromagnetic field.

\section{The spinor connection}

Not surprisingly, various complications arise when it comes to gravity. A chief reason for this is the 
fact that the gauge group for gravity $SO(1,3)$ (dealt with by means of the spin-1/2 spinor 
representation) is nonabelian unlike $U(1)$, and that the matrix representatives of its elements do not commute with the Dirac matrices. However, the essential features of the derivation in the last section undergo no change. We once again split the Lagrangian density
\begin{equation}
\mathcal{L} = \mathcal{L}_{1/2} + \mathcal{L}_{G}
\end{equation}
with $\mathcal{L}_{G}$ being the field Lagrangian density whose form is to be determined and $\mathcal{L}_{1/2}$ being the particle Lagrangian density given by
\begin{equation}
\mathcal{L}_{1/2} = \frac{i}{2} \left[\bar\psi\gamma^\mu(\psi_{,\mu} + \Gamma_\mu\psi) - (\bar\psi_{,\mu}+\bar\psi \bar\Gamma_\mu)\gamma^\mu\psi \right] - m \bar\psi \psi
\end{equation}
Here, $\Gamma_\mu$ is the gravitational connection analogous to the electromagnetic potential $A_\mu$. 
As it would be expected of a connection form associated with a principal $G$-bundle, it takes values in the Lie algebra $\mathfrak{g}$ associated with $G$. In the case of gravity, $G$ is the Lorentz group $SO(1,3)$, and its Lie algebra $\mathfrak{so}(1,3)$ carries a matrix representation wherein the space of matrices is spanned by $\sigma_{AB} =\frac{i}{2}\left[\gamma_A,\gamma_B\right]$, when the matrices $\gamma_A$ being the 
metric-independent Dirac matrices (the indices are denoted with the capital Latin alphabet to 
denote that they are internal).
\\

Next, we perform the following gauge transformations: $\psi\rightarrow\psi^\prime$, $\bar\psi\rightarrow\bar\psi^\prime$, $\gamma^\mu\rightarrow\gamma^{\prime\mu}$, $\Gamma_\mu\rightarrow\Gamma_\mu^\prime$ and $\bar\Gamma_\mu\rightarrow\bar\Gamma_\mu^\prime$ where
\begin{subequations}
\begin{align}
\psi^\prime &= L\psi\\
\bar\psi^\prime &= \bar\psi L^{-1}\\
\gamma^{\prime\mu} &= L\gamma^\mu L^{-1}
\end{align}
\end{subequations}
The last two transformations are to be determined.
\\

The matrix $L$ is a spin-1/2 representative of a Lorentz transformation, given by $\exp\left(i\epsilon^{AB}\sigma_{AB}\right)$, where $\epsilon^{AB}$ is the generator of the Lorentz transformation in its rank 2 tensor representation.
\\

The result of the above transformations leads to
\begin{align}
\mathcal{L}_{1/2}^\prime &= \frac{i}{2} \left[\bar\psi^\prime\gamma^{\prime\mu}(\psi^\prime_{,\mu} + \Gamma^\prime_\mu\psi^\prime) - (\bar\psi^\prime_{,\mu}+\bar\psi^\prime \bar\Gamma_\mu^\prime)\gamma^{\prime\mu}\psi^\prime \right] - m \bar\psi^\prime \psi^\prime\notag\\
&= \frac{i}{2} \left[\bar{\psi} L^{-1}L\gamma^\mu L^{-1}((L\psi)_{,\mu} + \Gamma^\prime_\mu L\psi)\right.\notag\\
  &\qquad -\left.((\bar{\psi} L^{-1})_{,\mu}+\bar{\psi} L^{-1} \bar\Gamma_\mu^\prime)L\gamma^\mu L^{-1} L\psi \right] - m \bar{\psi} L^{-1} L\psi\notag\\
&= \frac{i}{2} \left[\bar{\psi}\gamma^\mu(L^{-1}L_{,\mu}\psi +\psi_{,\mu} + L^{-1}\Gamma^\prime_\mu L\psi)\right.\notag\\
  &\qquad -\left.(\bar{\psi}_{,\mu}  + \bar{\psi} L^{-1}_{,\mu}L+\bar{\psi} L^{-1} \bar\Gamma_\mu^\prime L)\gamma^\mu \psi \right] - m \bar{\psi} \psi
\end{align}
Since, the Lagrangian density $\mathcal{L}_{1/2}$ is required to be invariant (i.e.~$\mathcal{L}_{1/2}=\mathcal{L}_{1/2}^\prime$), we have the following transformation laws:
\begin{subequations}
\begin{align}
\Gamma_\mu^\prime &= L\Gamma_\mu L^{-1} -L_{,\mu}L^{-1}\\
\bar\Gamma_\mu^\prime &= L\bar\Gamma_\mu L^{-1} -LL^{-1}_{,\mu}
\end{align}
\end{subequations}
Furthermore, as we have $L_{,\mu}L^{-1}+L L^{-1}_{,\mu} = \left(L L^{-1}\right)_{,\mu} = 1_{,\mu} = 0$, we note that $\bar\Gamma_\mu$ transforms exactly like $-\Gamma_\mu$. Since the spinor connection is defined by its transformation law alone (any information not deducible from that is a question of representation), we set
\begin{equation}
\bar\Gamma_\mu = -\Gamma_\mu
\end{equation}
Therefore, all we require is
\begin{equation}\label{gravconnectiontransform}
\Gamma_\mu^\prime = L\Gamma_\mu L^{-1} +LL^{-1}_{,\mu}
\end{equation}
We have already remarked that the spinor connection is an element of the Lie algebra spanned by $\sigma_{AB}$. Thus, $\Gamma_\mu$ may be represented as a linear combination of $\sigma_{AB}$
\begin{equation}
\Gamma_\mu = \frac{i}{4}\omega^{AB}_\mu\sigma_{AB}
\end{equation}
where the factor $i/4$ has been introduced to maintain notational consistency with Kaku \cite{Kaku} and the coefficients $\omega^{AB}_\mu$ are referred to as the Fock-Ivanenko coefficients (which may be taken to be antisymmetric in indices $A$ and $B$ as they are contracted with $A$ and $B$ in $\sigma_{AB}$ which is antisymmetric in $A$ and $B$).
\\

As it is only the Fock-Ivanenko coefficients which undergo changes under gauge transformations 
and which exhibit coordinate dependence, rather than the matrices $\sigma_{AB}$, it is clear that all gravitational dynamics will be encoded in them. Thus, our next step will be to construct an expression in terms of these coefficients that is invariant under gauge transformations.

\section{The gauge curvature}

Motivated by our analysis in Section 2, we begin by examining the gauge curvature which in the present case is given by
\begin{equation}
\Omega_{\mu\nu} = \Gamma_{\nu,\mu}-\Gamma_{\mu,\nu}+[\Gamma_\mu,\Gamma_\nu]
\end{equation}
To see how it transforms, we replace all the quantities above with their primed counterparts and use the transformation law \eqref{gravconnectiontransform} given by
\begin{align}
\Omega_{\mu\nu}^\prime &= \Gamma^\prime_{\nu,\mu}-\Gamma^\prime_{\mu,\nu}+[\Gamma^\prime_\mu,\Gamma^\prime_\nu]\notag\\
&= \left(L\Gamma_\nu L^{-1}\right)_{,\mu} +\left(LL^{-1}_{,\nu}\right)_{,\mu}-\left(L\Gamma_\mu L^{-1}\right)_{,\nu} -\left(LL^{-1}_{,\mu}\right)_{,\nu}\notag\\
&\qquad +~[L\Gamma_\mu L^{-1} +LL^{-1}_{,\mu},L\Gamma_\nu L^{-1} +LL^{-1}_{,\nu}]\notag\\
&= L\Gamma_{\nu,\mu}L^{-1}-L\Gamma_{\mu,\nu}L^{-1}+L[\Gamma_\mu,\Gamma_\nu]L^{-1}\notag\\
&= L\Omega_{\mu\nu}L^{-1}\label{gravgaugetransform}
\end{align}
The gauge curvature may be written in terms of the Fock-Ivanenko coefficients. 
\begin{align}
\Omega_{\mu\nu} &= \frac{i}{4}\omega^{AB}_{\nu,\mu}\sigma_{AB}-\frac{i}{4}\omega^{AB}_{\mu,\nu}\sigma_{AB}+\left[\frac{i}{4}\omega^{IJ}_{\mu}\sigma_{IJ},\frac{i}{4}\omega^{KL}_{\nu}\sigma_{KL}\right]\notag\\
&= \frac{i}{4}\left[\omega^{AB}_{\nu,\mu}\sigma_{AB}-\omega^{AB}_{\mu,\nu}\sigma_{AB}+\frac{i}{4}\omega^{IJ}_{\mu}\omega^{KL}_{\nu}\left[\sigma_{IJ},\sigma_{KL}\right]\right]\notag\\
&= \frac{i}{4}\left[\omega^{AB}_{\nu,\mu}\sigma_{AB}-\omega^{AB}_{\mu,\nu}\sigma_{AB}+\frac{i}{8}\left(\omega^{IJ}_{\mu}\omega^{KL}_{\nu}-\omega^{IJ}_{\nu}\omega^{KL}_{\mu}\right)\left[\sigma_{IJ},\sigma_{KL}\right]\right]\notag\\
&= \frac{i}{4}\left[\omega^{AB}_{\nu,\mu}-\omega^{AB}_{\mu,\nu}-\frac{1}{4}\left(\omega^{IJ}_{\mu}\omega^{KL}_{\nu}-\omega^{IJ}_{\nu}\omega^{KL}_{\mu}\right)\delta^{PA}_{IJ}\delta^{QB}_{KL}\eta_{PQ}\right]\sigma_{AB}
\end{align}
where $\eta_{PQ}$ is the usual Minkowski metric and $\delta^{PA}_{IJ}$ is the generalised Kronecker delta given by
\begin{equation*}
\delta^{PA}_{IJ}=\delta^P_I\delta^A_J-\delta^P_J\delta^A_I
\end{equation*}
The proof of the fact
\begin{equation*}
\left[\sigma_{IJ},\sigma_{KL}\right] = -2i\delta^{PA}_{IJ}\delta^{QB}_{KL}\eta_{PQ}\sigma_{AB}
\end{equation*}
used above is given in the Appendix.
\\

If we let
\begin{equation}
R^{AB}_{\mu\nu}=\frac{1}{4}\left[\omega^{AB}_{\nu,\mu}-\omega^{AB}_{\mu,\nu}-\frac{1}{4}\left(\omega^{IJ}_{\mu}\omega^{KL}_{\nu}-\omega^{IJ}_{\nu}\omega^{KL}_{\mu}\right)\delta^{PA}_{IJ}\delta^{QB}_{KL}\eta_{PQ}\right]
\end{equation}
the gauge curvature may be succinctly written as
\begin{equation}
\Omega_{\mu\nu} = iR^{AB}_{\mu\nu}\sigma_{AB}
\end{equation}
It may be easily verified that $R^{AB}_{\mu\nu}$ is antisymmetric in $A$ and $B$.
\\

Now, under a gauge transformation, elements in the Lie algebra associated with the gauge group get mapped to other elements in the said Lie algebra. In particular, $\Omega_{\mu\nu}$ gets mapped to $\Omega^\prime_{\mu\nu}$. The basis matrices $\sigma_{AB}$, $A<B$, prior to the gauge transformation, however, continue to serve as a basis for the transformed elements. It is therefore a matter of choice that we will preserve the Dirac commutators $\sigma_{AB}$ as they are and let only the coefficients transform. Thus, we have
\begin{equation}
\Omega_{\mu\nu}^\prime = iR^{\prime AB}_{\mu\nu}\sigma_{AB}
\end{equation}
Substituting the above into \eqref{gravgaugetransform} we obtain
\begin{align}
R^{\prime AB}_{\mu\nu}\sigma_{AB} &= LR^{UV}_{\mu\nu}\sigma_{UV}L^{-1}\notag\\
&= \frac{i}{2}R^{UV}_{\mu\nu}\left[L\gamma_UL^{-1},L\gamma_VL^{-1}\right]\notag\\
&= \frac{i}{2}R^{UV}_{\mu\nu}\left[\Lambda^A_{~U}\gamma_A,\Lambda^B_{~V}\gamma_B\right]\notag\\
&= \Lambda^A_{~U}\Lambda^B_{~V}R^{UV}_{\mu\nu}\sigma_{AB}
\end{align}
where $\Lambda^A_{~U}$ is the rank 2 tensor representative of the Lorentz transformation denoted by $L$ and the third line follows from the second by the virtue of the form-invariance of the Dirac equation in flat spacetime. If we shuffle the above onto one side we obtain
\begin{equation*}
\left(R^{\prime AB}_{\mu\nu}-\Lambda^A_{~U}\Lambda^B_{~V}R^{UV}_{\mu\nu}\right)\sigma_{AB}=0
\end{equation*}
We may use the fact that all the matrices $\sigma_{AB}$ with $A<B$ are linearly independent to assert. Hence
\begin{equation}
R^{\prime AB}_{\mu\nu}=\Lambda^A_{~U}\Lambda^B_{~V}R^{UV}_{\mu\nu}
\end{equation}
This is exactly how a contravariant Lorentz tensor of rank $2$ would transform. 
\\

The quantity $R^{AB}_{\mu\nu}$ is the exact analogue of the electromagnetic field tensor $F_{\mu\nu}$ -- it is, in addition to being antisymmetric in the indices $A$ and $B$ as we had observed earlier, also antisymmetric in $\mu$ and $\nu$. Motivated by our observations in Section 2, we might consider $\eta_{AM}\eta_{BN}g^{\mu\lambda}g^{\nu\rho}R^{AB}_{\mu\nu}R^{MN}_{\lambda\rho}$ as a probable choice for a gauge-invariant diffeomorphism-invariant scalar. But again, empirical considerations (more precisely, the fact that in the weak field Newtonian limit, the $1/r^2$ decay of the gravitational `force' holds across all scales \cite{Weinberg}) dictate that we reduce the number of contractions as far as possible while ensuring that our scalar does not identically vanish. In this case, owing to the presence of two internal indices $A$ and $B$ in addition to the spacetime indices $\mu$ and $\nu$, we can do so at the expense of introducing a new independent field of linear maps $e^\mu_A$ that takes basis vectors in the internal space of symmetry associated with a point to coordinate basis vectors dwelling in the tangent space at the point. By construction, such maps would behave as Lorentz 1-forms under gauge transformations and as vectors under general coordinate transformations. Therefore, we take $e^\mu_Ae^\nu_B R^{AB}_{\mu\nu}$ as the required gauge-invariant diffeomorphism-invariant scalar and let the field Lagrangian density for gravity be
\begin{equation}
\mathcal{L}_G = Ke^\mu_Ae^\nu_BR^{ AB}_{\mu\nu}
\end{equation}
where $K$ is a constant of proportionality.
\\

While we ought to be wary about introducing new independent fields leading to new degrees of freedom, 
we have not explicitly put in geometric information by hand -- no assumptions have been made regarding the possible dependence of $e^\mu_A$ on the metric. Hence, as long as we treat these fields as just `scaffolds' of sort that carry no physical meaning \emph{a priori}, our construction is a legitimate one. 

\section{The Palatini action}
 
The action associated with the Lagrangian density of spinors in presence of gravitational
interaction we have arrived at is thus
\begin{align}
S &= \int \mathcal{L} \sqrt{-g}~d^4x\notag\\
&= \int \left(\mathcal{L}_{1/2}+\mathcal{L}_{G}\right)\sqrt{-g}~d^4x\notag\\
&= \int \left(\frac{i}{2} \left[\bar\psi\gamma^\mu(\psi_{,\mu} +\Gamma_\mu\psi) - (\bar\psi_{,\mu}-\bar\psi \Gamma_\mu)\gamma^\mu\psi \right]\right.\notag\\
&\qquad \left.\indent\indent\indent-~m \bar\psi \psi+Ke^\mu_Ae^\nu_BR^{AB}_{\mu\nu}\right)\sqrt{-g}~d^4x
\end{align}
This is known as the Palatini action \cite{Baez} (strictly speaking, it is known as the Palatini action once the maps $e^\mu_A$ have been identified with the vierbein fields; that is an issue that will be addressed later). On carrying out a variation about its stationary points with respect to $\psi$, $\bar\psi$, $\omega^{AB}_{\nu}$ and $e^\nu_A$, we obtain the Euler-Lagrange equations
\begin{subequations}
\begin{align}
\partial_\mu\left(\frac{\partial\left(\mathcal{L}\sqrt{-g}\right)}{\partial\psi_{,\mu}}\right) - \frac{\partial\left(\mathcal{L}\sqrt{-g}\right)}{\partial\psi} &= 0\\
\partial_\mu\left(\frac{\partial\left(\mathcal{L}\sqrt{-g}\right)}{\partial\bar\psi_{,\mu}}\right) - \frac{\partial\left(\mathcal{L}\sqrt{-g}\right)}{\partial\bar\psi} &= 0\\
\partial_\mu\left(\frac{\partial\left(\mathcal{L}\sqrt{-g}\right)}{\partial \omega^{AB}_{\nu,\mu}}\right) - \frac{\partial\left(\mathcal{L}\sqrt{-g}\right)}{\partial \omega^{AB}_{\nu}} &= 0\\
\partial_\mu\left(\frac{\partial\left(\mathcal{L}\sqrt{-g}\right)}{\partial e^\nu_{A,\mu}}\right) - \frac{\partial\left(\mathcal{L}\sqrt{-g}\right)}{\partial e^\nu_{A}} &= 0\label{EL4}
\end{align}
\end{subequations}
Again using the fact that $\left(\partial_\mu\sqrt{-g}\right)=\frac{1}{2}\sqrt{-g}g^{\lambda\nu}g_{\lambda\nu,\mu}$, we may factor out $\sqrt{-g}$ to obtain
\begin{subequations}\label{ELeq}
\begin{align}
\frac{1}{2}g^{\lambda\rho}g_{\lambda\rho,\mu}\left(\frac{\partial\mathcal{L}}{\partial\psi_{,\mu}}\right)+\partial_\mu\left(\frac{\partial\mathcal{L}}{\partial\psi_{,\mu}}\right) - \frac{\partial\mathcal{L}}{\partial\psi} &= 0\\
\frac{1}{2}g^{\lambda\rho}g_{\lambda\rho,\mu}\left(\frac{\partial\mathcal{L}}{\partial\bar\psi_{,\mu}}\right)+\partial_\mu\left(\frac{\partial\mathcal{L}}{\partial\bar\psi_{,\mu}}\right) - \frac{\partial\mathcal{L}}{\partial\bar\psi} &= 0\\
\frac{1}{2}g^{\lambda\rho}g_{\lambda\rho,\mu}\left(\frac{\partial\mathcal{L}}{\partial \omega^{AB}_{\nu,\mu}}\right)+\partial_\mu\left(\frac{\partial\mathcal{L}}{\partial \omega^{AB}_{\nu,\mu}}\right) - \frac{\partial\mathcal{L}}{\partial \omega^{AB}_{\nu}} &= 0\\
\frac{1}{2}g^{\lambda\rho}g_{\lambda\rho,\mu}\left(\frac{\partial\mathcal{L}}{\partial e^\nu_{A,\mu}}\right)+\partial_\mu\left(\frac{\partial\mathcal{L}}{\partial e^\nu_{A,\mu}}\right) - \frac{\partial\mathcal{L}}{\partial e^\nu_{A}} &= 0
\end{align}
\end{subequations}
We compute the derivatives
\begin{align*}
\frac{\partial\mathcal{L}}{\partial\psi} &= \frac{i}{2}\left[ \bar\psi\gamma^\mu \Gamma_\mu - (\bar\psi_{,\mu}-\bar\psi \Gamma_\mu)\gamma^\mu\right]  - m \bar\psi\\
\frac{\partial\mathcal{L}}{\partial\bar\psi} &= \frac{i}{2} \left[\gamma^\mu(\psi_{,\mu} +\Gamma_\mu\psi) +\Gamma_\mu\gamma^\mu\psi \right] - m\psi\\
\frac{\partial\mathcal{L}}{\partial \omega^{AB}_{\nu}}&=-\frac{1}{8}\bar\psi\left\{\gamma^\nu,\sigma_{AB}\right\}\psi -\frac{K}{16}e^\xi_Ue^\zeta_V\left(\delta^{IJ}_{AB}\delta^\nu_\xi\omega^{KL}_\zeta-\omega^{IJ}_\xi\delta^{KL}_{AB}\delta^\nu_\zeta \right.\\
&\qquad\left.-~\delta^{IJ}_{AB}\delta^\nu_\zeta\omega^{KL}_\xi+\omega^{IJ}_\zeta\delta^{KL}_{AB}\delta^\nu_\xi\right)\delta^{PU}_{IJ}\delta^{QV}_{KL}\eta_{PQ}\\
\frac{\partial\mathcal{L}}{\partial e^\nu_{A}}&= 2Ke^\xi_UR^{AU}_{\nu\xi}\\
\frac{\partial\mathcal{L}}{\partial\psi_{,\mu}}&=\frac{i}{2}\bar\psi\gamma^\mu \\
\frac{\partial\mathcal{L}}{\partial\bar\psi_{,\mu}}&=-\frac{i}{2}\gamma^\mu\psi \\
\frac{\partial\mathcal{L}}{\omega^{AB}_{\nu, \mu}} &= \frac{K}{4}e^\xi_Ue^\zeta_V\left(\delta^\mu_\xi\delta^\nu_\zeta-\delta^\mu_\zeta\delta^\nu_\xi\right)\delta^{UV}_{AB}\\
&= \frac{K}{2}\left(e^\mu_Ae^\nu_B -e^\nu_Ae^\mu_B\right)\\
\frac{\partial\mathcal{L}}{\partial e^\nu_{A,\mu}}&=0
\end{align*}
Plugging the above into the Euler-Lagrange equations (30) yields the following:
\begin{subequations}
\begin{align}
i\left(\bar\psi_{,\mu}\gamma^\mu-\frac{1}{2}\bar\psi\{\gamma^\mu, \Gamma_\mu\}\right) +m\bar\psi &= -\frac{i}{2}\left(\frac{1}{2}g^{\lambda\rho}g_{\lambda\rho,\mu}\bar\psi\gamma^\mu+\bar\psi\gamma^\mu_{~,\mu}\right)\\
i\left(\gamma^\mu\psi_{,\mu}+\frac{1}{2}\{\gamma^\mu, \Gamma_\mu\}\psi\right) -m\psi &= -\frac{i}{2}\left(\frac{1}{2}g^{\lambda\rho}g_{\lambda\rho,\mu}\gamma^\mu\psi+\gamma^\mu_{~,\mu}\psi\right)\\
\left.\begin{array}{rr}
K\left(\frac{1}{2}g^{\lambda\rho}g_{\lambda\rho,\mu}\left(e^\mu_Ae^\nu_B -e^\nu_Ae^\mu_B\right)+e^\mu_{A,\mu}e^\nu_B\right.\\
\left. -~e^\nu_{A,\mu}e^\mu_B+e^\mu_Ae^\nu_{B,\mu} -e^\nu_Ae^\mu_{B,\mu}\right)
\end{array}
\right\} &= \left\{
\begin{array}{ll}-\bar\psi\left\{\gamma^\nu,\sigma_{AB}\right\}\psi-\frac{K}{4}e^\xi_Ue^\zeta_V\left(\delta^{IJ}_{AB}\delta^\nu_\xi\omega^{KL}_\zeta -\omega^{IJ}_\xi\delta^{KL}_{AB}\delta^\nu_\zeta\right.\\
\left. -~\delta^{IJ}_{AB}\delta^\nu_\zeta\omega^{KL}_\xi+\omega^{IJ}_\zeta\delta^{KL}_{AB}\delta^\nu_\xi\right)\delta^{PU}_{IJ}\delta^{QV}_{KL}\eta_{PQ}
\end{array}\right.\\
e^\xi_UR^{AU}_{\nu\xi}&=0
\end{align}
\end{subequations}
The first two equations are again the adjoint Dirac and Dirac equations with gravitational connection. 
The third equation governs the dynamics of the connection field. The fourth equation, however, presents a problem -- it implies that $\mathcal{L}_G = Ke^\mu_Ae^\nu_BR^{ AB}_{\mu\nu}$ identically vanishes, which is not permissible. To address this issue we take a second look at the fourth Euler-Lagrange equation \eqref{EL4}.
By construction we had taken the fields $e^\nu_A$ (and their first derivatives) and $\omega^{AB}_{\nu}$ (and its first derivative) to be independent, which implies that the following must necessarily continue to hold
\begin{align*}
\frac{\partial\mathcal{L}}{\partial e^\nu_{A}}&= 2Ke^\xi_UR^{AU}_{\nu\xi}\\
\frac{\partial\mathcal{L}}{\partial e^\nu_{A,\mu}}&=0
\end{align*} 
Hence, the only remedy to ensure that $\mathcal{L}_G$ does not vanish is to stipulate that $\partial\sqrt{-g}/\partial e^\nu_{A}$ is nonzero. In other words, we introduce a dependence of the metric on the fields $ e^\nu_{A}$ which shall be henceforth referred to as a vierbein. The fourth Euler-Lagrange equation hence becomes
\begin{equation}
2Ke^\xi_UR^{AU}_{\nu\xi}+\frac{1}{2}\mathcal{L}g^{\lambda\rho}\frac{\partial g_{\lambda\rho}}{\partial e^\nu_{A}}=0
\end{equation}
The above equation carries information about the relationship between the vierbein and the metric as well as that between the vierbein and the matter fields. Since we are interested only in the former, we set the matter fields to zero i.e.~$\mathcal{L}=\mathcal{L}_G$. Furthermore contracting with $e^\nu_A$ throughout and using the fact that $g^{\lambda\rho}g_{\lambda\rho}=4$ which is a constant, we have
\begin{equation}
2K\mathcal{L}_G-\frac{1}{2}\mathcal{L}_Ge^\nu_{A}\frac{\partial g^{\lambda\rho}}{\partial e^\nu_{A}}g_{\lambda\rho}=0
\end{equation}
In other words, we are looking for solutions that satisfy the following differential equation for any metric.
\begin{equation}
e^\nu_{A}\frac{\partial g^{\lambda\rho}}{\partial e^\nu_{A}}g_{\lambda\rho}=4K
\end{equation}
The solutions turn out to be
\begin{equation}
g^{\xi\zeta}=4Ke^\xi_Ue^\zeta_V\eta^{UV}
\end{equation}
Since $e^\xi_U$ may be scaled howsoever we please, we take $4K$ to be 1, so that 
\begin{equation}
g^{\xi\zeta}=e^\xi_Ue^\zeta_V\eta^{UV}
\end{equation}
The Euler-Lagrange equations now become
\begin{subequations}
\begin{align}
i\left(\bar\psi_{,\mu}\gamma^\mu-\frac{1}{2}\bar\psi\{\gamma^\mu, \Gamma_\mu\}\right) +m\bar\psi &= -\frac{i}{2}\left(\frac{1}{2}g^{\lambda\rho}g_{\lambda\rho,\mu}\bar\psi\gamma^\mu+\bar\psi\gamma^\mu_{~,\mu}\right)\\
i\left(\gamma^\mu\psi_{,\mu}+\frac{1}{2}\{\gamma^\mu, \Gamma_\mu\}\psi\right) -m\psi &= -\frac{i}{2}\left(\frac{1}{2}g^{\lambda\rho}g_{\lambda\rho,\mu}\gamma^\mu\psi+\gamma^\mu_{~,\mu}\psi\right)\label{Diracgrav}\\
\left.\begin{array}{rr}
\frac{1}{4}\left(\frac{1}{2}g^{\lambda\rho}g_{\lambda\rho,\mu}\left(e^\mu_Ae^\nu_B -e^\nu_Ae^\mu_B\right)+e^\mu_{A,\mu}e^\nu_B\right.\\
\left. -~e^\nu_{A,\mu}e^\mu_B+e^\mu_Ae^\nu_{B,\mu} -e^\nu_Ae^\mu_{B,\mu}\right)
\end{array}
\right\} &= \left\{
\begin{array}{ll}-\bar\psi\left\{\gamma^\nu,\sigma_{AB}\right\}\psi-\frac{1}{16}e^\xi_Ue^\zeta_V\left(\delta^{IJ}_{AB}\delta^\nu_\xi\omega^{KL}_\zeta -\omega^{IJ}_\xi\delta^{KL}_{AB}\delta^\nu_\zeta\right.\\
\left. -~\delta^{IJ}_{AB}\delta^\nu_\zeta\omega^{KL}_\xi+\omega^{IJ}_\zeta\delta^{KL}_{AB}\delta^\nu_\xi\right)\delta^{PU}_{IJ}\delta^{QV}_{KL}\eta_{PQ}
\end{array}\right.\\
e^\xi_U\left(R^{AU}_{\nu\xi}-2\mathcal{L}g_{\nu\xi}\eta^{AU}\right)&=0
\end{align}
\end{subequations}
The first two equations have undergone no change. The third and fourth equations are essentially the Einstein-Cartan field equations in vierbein formalism. 
\section{Remarks on geometric content}
The vierbein fields introduced in Section 4 can thus be regarded as linear maps either from the space of Lorentz vectors in the internal space of symmetry associated with a point to the tangent space at the point or from the cotangent space at a point to the space of Lorentz 1-forms in the internal space of symmetry associated with the point. In particular, it maps the metric tensor $g_{\mu\nu}$ to the Minkowski metric $\eta_{AB}$ and, as a result, preserves inner products between vectors in the internal space of symmetry and the tangent space. Physically, this means that the vierbein is a local freely falling frame and contains all the information about the metric. Moreover, this was not the result of an arbitrary choice a priori but was demanded by the necessary condition that the field Lagrangian density be the simplest possible and nontrivial (by which we mean that the number of contractions is minimised and the Lagrangian density does not identically vanish). The connection to geometry thus arises as a consequence of the above observations, as we had claimed in the Introduction.
\\

However, a loose end persists. The connection field coefficients $\omega^{AB}_\mu$ are still independent of the vierbein and the metric. But its additional independent degrees of freedom actually offer an advantage over the Einstein field equations. In the original theory, the stress-energy-momentum tensor $T_{\mu\nu}$ was required to be symmetric in the indices $\mu$ and $\nu$. This means that it failed to account for the effect of spin-orbit coupling, whose contribution to the stress-energy-momentum tensor is nonsymmetric and which invariably is an issue if we are to talk of the `motion' of spin-1/2 particles in a gravitational field. The extra degrees of freedom in $\omega^{AB}_\mu$ constitutes the torsion which \emph{does} account for spin-orbit coupling \cite{Torsion}. The resulting theory i.e.~Einstein-Cartan gravity is therefore more general and powerful than Einstein's original.
\\

However, if the effects of spin-orbit coupling are neglected, as we shall in the rest of this paper, the torsion may be set to identically vanish, and the extra degrees of freedom in $\omega^{AB}_\mu$ is eliminated. In such a case, the connection would be related to the vierbein and the metric by \cite{BanibrataMukhopadhyay}
\begin{equation}
\omega^{AB}_\mu = e^A_\lambda\eta^{BI}(e^{\lambda}_{I,\mu}+\Gamma^\lambda_{\gamma \mu}e^{\gamma }_I)
\end{equation}
where $e^A_\lambda=g_{\lambda\rho}\eta^{AI}e^\rho_I$ is the inverse vierbein and $\Gamma^\lambda_{\gamma \mu}=\frac{1}{2} g^{\lambda\rho}\left(g_{\rho\gamma,\mu}+g_{\rho\mu,\gamma }-g_{\gamma \mu,\rho}\right)$ is the Christoffel symbol. In such a case, the gravitational field becomes a purely geometric entity.
\section{The Klein-Gordon equation in a gravitational field}
To obtain the Klein-Gordon equation in a gravitational field, we simply shuffle the term $m\psi$ in \eqref{Diracgrav} onto one side and the rest of the terms onto the other, and rewrite everything in the 
operator formalism as
\begin{equation}
i\left(\gamma^\mu\partial_\mu+ \frac{1}{2}\{\gamma^\mu, \Gamma_\mu\}+\frac{1}{4}g^{\lambda\rho}g_{\lambda\rho,\mu}\gamma^\mu + \frac{1}{2}\gamma^\mu_{,\mu}\right)\psi= -m\psi
\end{equation}
The rationale for isolating the $m\psi$ term above is that $m$ is a scalar and commutes with any operator, hence allowing us to apply the same operator twice without worrying about extra commutators turning up,
which yields
\begin{equation}
\left(\gamma^\mu\partial_\mu+ \frac{1}{2}\{\gamma^\mu, \Gamma_\mu\}+\frac{1}{4}g^{\lambda\rho}g_{\lambda\rho,\mu}\gamma^\mu + \frac{1}{2}\gamma^\mu_{,\mu}\right)^2\psi= -m^2\psi
\end{equation}
The distributivity of operator composition over operator sums allows the above to be rewritten in the form
\begin{equation}
\left(\partial^\mu \partial_\mu + U^\nu \partial_\nu + V\right)\psi = -m^2\psi
\end{equation}
where the coefficients $U^\nu$ and $V=V_{(1)}+V_{(2)}+V_{(3)}$ are given by
\begin{subequations}
\begin{align}
U^\nu &= \gamma^\mu\gamma^\nu_{,\mu}+ \frac{1}{2}\{\gamma^\mu_{,\mu},\gamma^\nu\}+ \frac{1}{2}\{\{\gamma^\mu, \Gamma_\mu\},\gamma^\nu\}+\frac{1}{2}g^{\lambda\rho}g_{\lambda\rho,\mu}g^{\mu\nu}\\
V_{(1)} &= \frac{1}{2}\gamma^\mu\{\gamma^\nu_{,\mu}, \Gamma_\nu\}+\frac{1}{2}\gamma^\mu\{\gamma^\nu, \Gamma_{\nu,\mu}\}+\frac{1}{4} g^{\lambda\rho}_{,\mu}g_{\lambda\rho,\nu}\gamma^\mu\gamma^\nu\notag\\
&\qquad +\frac{1}{4}g^{\lambda\rho}g_{\lambda\rho,\mu\nu}g^{\mu\nu}+\frac{1}{4}g^{\rho\lambda}g_{\lambda\rho,\nu}\gamma^\mu \gamma^\nu_{,\mu}+\frac{1}{2}\gamma^\mu\gamma^\nu_{,\mu\nu}\\
V_{(2)} &= \frac{1}{4}\{\gamma^\mu, \Gamma_\mu\}\{\gamma^\nu, \Gamma_\nu\}+\frac{1}{16}g^{\lambda\rho}g_{\lambda\rho,\mu}g^{\xi\zeta}g_{\xi\zeta,\nu}g^{\mu\nu}+\frac{1}{4}\gamma^\mu_{,\mu}\gamma^\nu_{,\nu}\\
V_{(3)} &= \frac{1}{8}g^{\lambda\rho}g_{\lambda\rho,\mu}\{\gamma^\mu,\{\gamma^\nu,\Gamma_\nu\}\} + \frac{1}{4}\{\gamma^\mu_{,\mu},\{\gamma^\nu,\Gamma_\nu\}\}+ \frac{1}{8}g^{\lambda\rho}g_{\lambda\rho,\mu}\{\gamma^\mu,\gamma^\nu_{,\nu}\}
\end{align}
\end{subequations}
On evaluating the anticommutators and simplifying them as far as possible, we have
\begin{subequations}
\begin{align}
U^\nu &= \gamma^\mu\gamma^\nu_{,\mu}+e^\mu_{I,\mu}e^\nu_{J}\eta^{IJ}+ \frac{i}{2}\omega^{IJ}_\mu e^\mu_K e^\nu_N\epsilon_{MIJ}^{~~~~~K}\left(\gamma^N\gamma^M-\eta^{MN}\right)\gamma^5+\frac{1}{2}g^{\lambda\rho}g_{\lambda\rho,\mu}g^{\mu\nu}\\
V_{(1)} &= \frac{i}{4}\left(\omega^{IJ}_\nu e^\nu_{K,\mu}+\omega^{IJ}_{\nu,\mu} e^\nu_{K}\right)\epsilon_{MIJ}^{~~~~~K}\gamma^\mu\gamma^M\gamma^5+\frac{1}{4} g^{\lambda\rho}_{,\mu}g_{\lambda\rho,\nu}\gamma^\mu\gamma^\nu\notag\\
&\qquad +\frac{1}{4}g^{\lambda\rho}g_{\lambda\rho,\mu\nu}g^{\mu\nu}+\frac{1}{4}g^{\rho\lambda}g_{\lambda\rho,\nu}\gamma^\mu \gamma^\nu_{,\mu}+\frac{1}{2}\gamma^\mu\gamma^\nu_{,\mu\nu}\\
V_{(2)} &= \frac{1}{16}\omega^{IJ}_\mu\omega^{PQ}_\nu e^\mu_K e^\nu_R\epsilon_{MIJ}^{~~~~~K} \epsilon_{NPQ}^{~~~~~R}\eta^{MN}+\frac{1}{16}g^{\lambda\rho}g_{\lambda\rho,\mu}g^{\xi\zeta}g_{\xi\zeta,\nu}g^{\mu\nu}+\frac{1}{4}e^\mu_{I,\mu}e^\nu_{J,\nu}\eta^{IJ}\\
V_{(3)} &=\frac{i}{8}g^{\lambda\rho}g_{\lambda\rho,\mu}\omega^{IJ}_\nu e^\nu_K e^\mu_N\epsilon_{MIJ}^{~~~~~K}\left(\gamma^N\gamma^M-\eta^{MN}\right)\gamma^5\notag\\
&\qquad +\frac{i}{4}\omega^{IJ}_\nu e^\nu_K e^\mu_{N,\mu}\epsilon_{MIJ}^{~~~~~K}\left(\gamma^N\gamma^M-\eta^{MN}\right)\gamma^5 + \frac{1}{4}g^{\lambda\rho}g_{\lambda\rho,\mu}e^\mu_{I}e^\nu_{J,\nu}\eta^{IJ}
\end{align}
\end{subequations}
where $\epsilon_{MIJK}$ is the Levi-Civita tensor.
\section{The weak field limit}
In the weak field limit, we assume the following to hold
\begin{subequations}
\begin{align}
g_{\mu\nu}&=\eta_{\bar\mu\bar\nu}+\varepsilon \Delta g_{\mu\nu}(x)+\mathcal{O}(\varepsilon^2)\\
g^{\mu\nu}&=\eta^{\bar\mu\bar\nu}-\varepsilon\eta^{\bar\mu\bar\xi}\eta^{\bar\nu\bar\zeta} \Delta g_{\xi\zeta}(x)+\mathcal{O}(\varepsilon^2)\\
e^\mu_I&=\delta^{\bar\mu}_I-\frac{1}{2}\varepsilon \eta^{\bar\mu\bar\xi}\delta^{\bar\zeta}_I \Delta g_{\xi\zeta}(x)+\mathcal{O}(\varepsilon^2)\\
e_\mu^I&=\delta_{\bar\mu}^I+\frac{1}{2}\varepsilon \eta^{I\bar\zeta} \Delta g_{\mu\zeta}(x)+\mathcal{O}(\varepsilon^2)\\
\omega^{IJ}_{\mu}&=\varepsilon \Delta\omega^{IJ}_{\mu}(x)+\mathcal{O}(\varepsilon^2)\notag\\
&=\frac{1}{2}\varepsilon\eta^{I\bar\xi}\eta^{J\bar\zeta}(\Delta g_{\mu\xi,\zeta}-\Delta g_{\mu\zeta,\xi})+\mathcal{O}(\varepsilon^2)
\end{align}
\end{subequations}
where the Greek indices with bar denote that they are to be considered on the same footing as of the 
capital Latin indices, $\varepsilon$ is an independent real parameter such that $\varepsilon\ll 1$ and the
$\Delta$ in $\Delta g_{\mu\nu}$ and $\Delta\omega^{IJ}_{\mu}$ implies a small variation.
\\

The coordinate-dependent Dirac matrices thus become
\begin{align}
\gamma^\mu &= e^\mu_I\gamma^I\notag\\
&=\delta^{\bar\mu}_I\gamma^I-\frac{1}{2}\varepsilon \eta^{\bar\mu\bar\xi}\delta^{\bar\zeta}_I \Delta g_{\xi\zeta}\gamma^I+\mathcal{O}(\varepsilon^2)\notag\\
&= \gamma^{\bar\mu}-\frac{1}{2}\varepsilon  \eta^{\bar\mu\bar\xi}\gamma^{\bar\zeta} \Delta g_{\xi\zeta}+\mathcal{O}(\varepsilon^2)
\end{align}
Therefore, to the first order in $\varepsilon$ we have
\begin{subequations}
\begin{align}
U^\nu&=-\frac{1}{2}\varepsilon\eta^{\bar\nu\bar\xi}\left(\gamma^{\bar\mu}\gamma^{\bar\zeta} + \eta^{\bar\mu\bar\zeta}\right)\Delta g_{\xi\zeta,\mu}+ \frac{i}{2}\varepsilon\Delta\omega^{IJ}_\mu \epsilon_{MIJ}^{~~~~~\bar\mu}\left(\gamma^{\bar\nu}\gamma^M-\eta^{M\bar\nu}\right)\gamma^5+\frac{1}{2}\varepsilon\eta^{\bar\lambda\bar\rho}\eta^{\bar\mu\bar\nu}\Delta g_{\lambda\rho,\mu}\notag\\
&= \frac{1}{2}\varepsilon\left[\left(-\eta^{\bar\nu\bar\xi}\left(\gamma^{\bar\mu}\gamma^{\bar\zeta} + \eta^{\bar\mu\bar\zeta}\right)+\eta^{\bar\xi\bar\zeta}\right)\Delta g_{\xi\zeta,\mu}+i\eta^{I\bar\xi}\eta^{J\bar\zeta}\Delta g_{\mu\xi,\zeta} \epsilon_{MIJ}^{~~~~~\bar\mu}\left(\gamma^{\bar\nu}\gamma^M-\eta^{M\bar\nu}\right)\gamma^5\right]\notag\\
&= \frac{1}{2}\varepsilon\Delta g_{\xi\zeta,\mu}\left[-\eta^{\bar\nu\bar\xi}\left(\gamma^{\bar\mu}\gamma^{\bar\zeta} + \eta^{\bar\mu\bar\zeta}\right)+\eta^{\bar\xi\bar\zeta}\eta^{\bar\mu\bar\nu}+i \epsilon_{M}^{~~\bar\xi\bar\mu\bar\zeta}\left(\gamma^{\bar\nu}\gamma^M-\eta^{M\bar\nu}\right)\gamma^5\right]\notag\\
&= \frac{1}{2}\varepsilon\Delta g_{\xi\zeta,\mu}\left[-\eta^{\bar\nu\bar\xi}\left(\gamma^{\bar\mu}\gamma^{\bar\zeta} + \eta^{\bar\mu\bar\zeta}\right)+\eta^{\bar\xi\bar\zeta}\eta^{\bar\mu\bar\nu}\right]\\
V &= \frac{i}{4}\varepsilon\Delta\omega^{IJ}_{\nu,\mu} \epsilon_{MIJ}^{~~~~~\bar\nu}\gamma^{\bar\mu}\gamma^M\gamma^5+\frac{1}{4}\varepsilon\left(\eta^{\bar\xi\bar\zeta}\eta^{\bar\mu\bar\nu}-\eta^{\bar\nu\bar\xi}\gamma^{\bar\mu}\gamma^{\bar\zeta}\right) \Delta g_{\xi\zeta,\mu\nu}\notag\\
&= \frac{1}{4}\varepsilon\left[\left(\eta^{\bar\xi\bar\zeta}\eta^{\bar\mu\bar\nu}-\eta^{\bar\nu\bar\xi}\gamma^{\bar\mu}\gamma^{\bar\zeta}\right) \Delta g_{\xi\zeta,\mu\nu}+i\eta^{I\bar\xi}\eta^{J\bar\zeta}\Delta g_{\nu\xi,\mu\zeta} \epsilon_{MIJ}^{~~~~~\bar\nu}\gamma^{\bar\mu}\gamma^M\gamma^5\right]\notag\\
&= \frac{1}{4}\varepsilon\Delta g_{\xi\zeta,\mu\nu}\left(\eta^{\bar\xi\bar\zeta}\eta^{\bar\mu\bar\nu}-\eta^{\bar\nu\bar\xi}\gamma^{\bar\mu}\gamma^{\bar\zeta} +i\epsilon_{M}^{~~\bar\xi\bar\nu\bar\zeta}\gamma^{\bar\mu}\gamma^M\gamma^5\right)\notag\\
&= \frac{1}{4}\varepsilon\Delta g_{\xi\zeta,\mu\nu}\left(\eta^{\bar\xi\bar\zeta}\eta^{\bar\mu\bar\nu}-\eta^{\bar\nu\bar\xi}\gamma^{\bar\mu}\gamma^{\bar\zeta}\right)
\end{align}
\end{subequations}
where the final steps follow from the fact that $\Delta g_{\xi\zeta}$ is symmetric in the indices $\xi$ and $\zeta$ while $\epsilon_{M}^{~~\bar\xi\bar\mu\bar\zeta}$ and $\epsilon_{M}^{~~\bar\xi\bar\nu\bar\zeta}$ are antisymmetric in them.

\section{The Schr\"odinger limit}
In this section, we shall be including the speed of light $c$ in our equations explicitly as we will have to the take the limit $1/c^2\ll1$. Accordingly, the operator $\partial_0$ shall be written as $\frac{1}{c}\frac{\partial}{\partial t}$. The Klein-Gordon equation hence becomes
\begin{equation}
\left(g^{00}\frac{1}{c^2}\frac{\partial^2}{\partial t^2}+ \frac{2}{c}\vec g\cdot\vec\nabla\frac{\partial}{\partial t}- \nabla^2 + \frac{1}{c}U^0 \frac{\partial}{\partial t}-\vec U\cdot\vec\nabla + V\right)\psi = -m^2\psi
\end{equation}
where $\vec g\cdot\vec\nabla$ denotes $g^{j0}\partial_j$, small letter Latin indices being understood to 
run over spatial indices (note that we have adopted the mostly minus convention). We shuffle the terms about and rewrite the above as
\begin{equation}
\left(-g^{00}\frac{1}{c^2}\frac{\partial^2}{\partial t^2}-\frac{1}{c}U^0 \frac{\partial}{\partial t}\right)\psi = \left(m^2+\frac{2}{c}\vec g\cdot\vec\nabla\frac{\partial}{\partial t}- \nabla^2 -\vec U\cdot\vec\nabla + V\right)\psi
\end{equation}
On completing the `squares' and multiplying by $c^2$ throughout, we have
\begin{align}
-\left(\sqrt{g^{00}}\frac{\partial}{\partial t} + \frac{cU^0-\dot{g}^{00}/2}{2\sqrt{g^{00}}}\right)^2\psi &= \left[m^2c^4-c^2\left(\vec\nabla+\frac{\vec U}{2}\right)^2 + \frac{c^2}{2}(\mathrm{div}~\vec{U})+\frac{c^2}{4}U^2+c^2V\right.\notag\\
&\qquad \left. -\frac{(c\dot U^0-\ddot{g}^{00}/2)g^{00}-(cU^0-\dot{g}^{00}/2)\dot g^{00}}{2g^{00}}+2c\vec g\cdot\vec\nabla\frac{\partial}{\partial t}\right]\psi\notag\\
&= m^2c^4\left[1-\frac{1}{m^2c^2}\left(\vec\nabla+\frac{\vec U}{2}\right)^2 + \frac{\mathrm{div}~\vec{U}}{2m^2c^2}+\frac{U^2}{4m^2c^2}+\frac{V}{m^2c^2}\right.\notag\\
&\qquad \left. -\frac{1}{m^2c^4}\frac{(c\dot U^0-\ddot{g}^{00}/2)g^{00}-(cU^0-\dot{g}^{00}/2)\dot g^{00}}{2g^{00}}+\frac{2}{m^2c^3}\vec g\cdot\vec\nabla\frac{\partial}{\partial t}\right]\psi
\end{align}
where $\dot f$ denotes the derivative of a function $f$ with respect to time. On taking the square roots of the operators on either side and expanding the right hand side to order $1/c^2$ in its binomial expansion, we have
\begin{align}
i\left(\sqrt{g^{00}}\frac{\partial}{\partial t} + \frac{cU^0-\dot{g}^{00}/2}{2\sqrt{g^{00}}}\right)\psi &= mc^2\left[1-\frac{1}{m^2c^2}\left(\vec\nabla+\frac{\vec U}{2}\right)^2 + \frac{\mathrm{div}~\vec{U}}{2m^2c^2}+\frac{U^2}{4m^2c^2}+\frac{V}{m^2c^2}\right.\notag\\
&\qquad \left. -\frac{1}{m^2c^4}\frac{(c\dot U^0-\ddot{g}^{00}/2)g^{00}-(cU^0-\dot{g}^{00}/2)\dot g^{00}}{2g^{00}}+\frac{2}{m^2c^3}\vec g\cdot\vec\nabla\frac{\partial}{\partial t}\right]^\frac{1}{2}\psi\notag\\
&\approx mc^2\left[1-\frac{1}{2m^2c^2}\left(\left(\vec\nabla+\frac{\vec U}{2}\right)^2 - \frac{\mathrm{div}~\vec{U}}{2}-\frac{U^2}{4}-V\right)\right]\psi\notag\\
&= \left[mc^2-\frac{1}{2m}\left(\left(\vec\nabla+\frac{\vec U}{2}\right)^2 - \frac{\mathrm{div}~\vec{U}}{2}-\frac{U^2}{4}-V\right)\right]\psi
\end{align}
Therefore, the Schr\"odinger equation for a (slowly moving) particle in a gravitational field is
\begin{equation}
i\sqrt{g^{00}}\frac{\partial \psi}{\partial t} = \left[mc^2-\frac{1}{2m}\left(\vec\nabla+\frac{\vec U}{2}\right)^2 + \frac{1}{2m}\left(\frac{\mathrm{div}~\vec{U}}{2}+\frac{U^2}{4}+V\right)-i\left(\frac{cU^0-\dot{g}^{00}/2}{2\sqrt{g^{00}}}\right)\right]\psi
\end{equation}
On comparing the above with the Schr\"odinger equation of a particle with charge $q$ in an electromagnetic field (without the Stern-Gerlach correction) 
\begin{equation*}
i\frac{\partial \psi}{\partial t} = \left[-\frac{1}{2m}\left(\vec\nabla -iq\vec A\right)^2+q\Phi\right]\psi
\end{equation*}
we see that the analogue of the magnetic potential $q\vec A$ is $\frac{i}{2}\vec{U}$ in gravity,
called `gravito-magnetic $3$-vector potential', and that of the electric potential $q\Phi$, 
once we have removed the constant term $mc^2$ which contributes only to a global change in the phase of 
the wavefunction $\psi$, is
$$
\frac{1}{2m}\left(\frac{\mathrm{div}~\vec{U}}{2}+\frac{U^2}{4}+V\right)-i\left(\frac{cU^0-\dot{g}^{00}/2}{2\sqrt{g^{00}}}\right)
$$
called `gravito-electric scalar potential'.
\section{Investigating the `gravito-magnetic potential' $\frac{i}{2}\vec{U}$}
We shall now be studying the real part of the `gravito-magnetic potential' $\frac{i}{2}\vec{U}$ given by
\begin{equation}
\mathrm{Re}\left(\frac{i}{2}\vec{U}\right)=-\frac{1}{4}\omega^{IJ}_\mu e^\mu_K e^\nu_N\epsilon_{MIJ}^{~~~~~K}\left(\gamma^N\gamma^M-\eta^{MN}\right)\gamma^5
\end{equation}
 in two different situations -- first, in terms of the Fermi normal coordinates, wherein the Christoffel 
symbols along a chosen geodesic vanish, and second, on the equatorial plane in the Kerr geometry.
Similarly, one can look at the real part of the gravito-electric potential. These, being Dirac self-adjoint, are the physical
observables.
\subsection{Fermi normal coordinates}
In the following, the lowercase Latin indices denote spatial indices for both the local freely 
falling coordinates as well as the global coordinate indices. No attempt has been made to distinguish between the two cases as it leads to no significant confusion here. We will be working with the following vierbein \cite{BanibrataMukhopadhyay}
\begin{subequations}
\begin{align}
e^\alpha_0&=\delta^{\alpha}_0-\frac{1}{2}R^\alpha_{~l0m}X^lX^m\\
e^\alpha_{ j}&=\delta^{\alpha}_{ j}-\frac{1}{6}R^\alpha_{~ljm}X^lX^m
\end{align}
\end{subequations}
The Fock-Ivanenko coefficients are thus given by
\begin{subequations}
\begin{align}
\omega^{IJ}_a &= e^I_\lambda\eta^{JK}(e^{\lambda}_{K,a}+\Gamma^\lambda_{\gamma a}e^{\gamma }_K)\notag\\
&= -k(J)(\delta^{I}_{\lambda}-k(I)R_{\lambda l~m}^{~~I}X^lX^m)(R_{~a~n}^{\lambda~J}+R_{~n~a}^{\lambda~J})X^n\\
\omega^{IJ}_0 &= 0
\end{align}
\end{subequations}
where $k(0)=1/2$ and $k(j)=1/6$, $j$ denoting a spatial index.
\\

For the sake of brevity, we let $\mathfrak{F}^{KN}_{IJ}=-\frac{1}{4}\epsilon_{MIJ}^{~~~~~K}\left(\gamma^N\gamma^M-\eta^{MN}\right)\gamma^5$. Then, we have
\begin{equation}
e^\mu_K e^\nu_N\mathfrak{F}^{KN}_{IJ}= e^\mu_0 e^\nu_0\mathfrak{F}^{00}_{IJ}+e^\mu_0 e^\nu_j\mathfrak{F}^{0j}_{IJ}+e^\mu_i e^\nu_0\mathfrak{F}^{i0}_{IJ}+e^\mu_i e^\nu_j\mathfrak{F}^{ij}_{IJ}
\end{equation}
where each individual term is further given by
\begin{subequations}
\begin{align}
e^\mu_0 e^\nu_0\mathfrak{F}^{00}_{IJ}&=\delta^\mu_0\delta^\nu_0\mathfrak{F}^{00}_{IJ}-\frac{1}{2}\delta^\mu_0\mathfrak{F}^{0 0}_{IJ}R^\nu_{~l0m}X^lX^m-\frac{1}{2}\delta^\nu_0\mathfrak{F}^{00}_{IJ}R^\mu_{~l0m}X^lX^m+\frac{1}{4}\mathfrak{F}^{00}_{IJ}R^\mu_{~l0m}R^\nu_{~p0q}X^lX^mX^pX^q\\
e^\mu_0 e^\nu_j\mathfrak{F}^{0j}_{IJ}&=\delta^\mu_0\delta^\nu_j\mathfrak{F}^{0j}_{IJ}-\frac{1}{6}\delta^\mu_0\mathfrak{F}^{0 j}_{IJ}R^\nu_{~ljm}X^lX^m-\frac{1}{2}\delta^\nu_j\mathfrak{F}^{0j}_{IJ}R^\mu_{~l0m}X^lX^m+\frac{1}{12}\mathfrak{F}^{0j}_{IJ}R^\mu_{~l0m}R^\nu_{~pjq}X^lX^mX^pX^q\\
e^\mu_i e^\nu_0\mathfrak{F}^{i0}_{IJ}&=\delta^\mu_i\delta^\nu_0\mathfrak{F}^{i0}_{IJ}-\frac{1}{2}\delta^\mu_i\mathfrak{F}^{i 0}_{IJ}R^\nu_{~l0m}X^lX^m-\frac{1}{6}\delta^\nu_0\mathfrak{F}^{i0}_{IJ}R^\mu_{~lim}X^lX^m+\frac{1}{12}\mathfrak{F}^{i0}_{IJ}R^\mu_{~lim}R^\nu_{~pjq}X^lX^mX^pX^q\\
e^\mu_0 e^\nu_0\mathfrak{F}^{00}_{IJ}&=\delta^\mu_i\delta^\nu_j\mathfrak{F}^{ij}_{IJ}-\frac{1}{6}\delta^\mu_i\mathfrak{F}^{i j}_{IJ}R^\nu_{~ljm}X^lX^m-\frac{1}{6}\delta^\nu_j\mathfrak{F}^{i\nu}_{IJ}R^\mu_{~lim}X^lX^m+\frac{1}{36}\mathfrak{F}^{ij}_{IJ}R^\mu_{~lim}R^\nu_{~pjq}X^lX^mX^pX^q
\end{align}
\end{subequations}
We also let $\mathfrak{G}^{\mu\nu}_{IJ}=e^\mu_K e^\nu_N\mathfrak{F}^{KN}_{IJ}$. Then, the real part of 
the `gravito-magnetic potential' may be written as
\begin{align}
\mathrm{Re}\left(\frac{i}{2}\vec{U}\right)&=-\frac{1}{4}\omega^{IJ}_\mu e^\mu_K e^\nu_N\epsilon_{MIJ}^{~~~~~K}\left(\gamma^N\gamma^M-\eta^{MN}\right)\gamma^5\notag\\
&=\omega^{IJ}_\mu\mathfrak{G}^{\mu\nu}_{IJ}\notag\\
&= \omega^{IJ}_a\mathfrak{G}^{a\nu}_{IJ}\notag\\
&= \omega^{0j}_a\mathfrak{G}^{a\nu}_{0j}+\omega^{i0}_a\mathfrak{G}^{a\nu}_{i0}+\omega^{ij}_a\mathfrak{G}^{a\nu}_{ij}
\end{align}
where each individual term is given by
\begin{subequations}
\begin{align}
\omega^{0j}_a\mathfrak{G}^{a\nu}_{0j}&= -\frac{1}{6}(\delta^0_\lambda\mathfrak{G}^{a\nu}_{0 j}-\frac{1}{2}\mathfrak{G}^{a\nu}_{0j}R_{\lambda l~m}^{~~0}X^lX^m)(R_{~a~n}^{\lambda~j}+R_{~n~a}^{\lambda~j})X^n\\
\omega^{i0}_a\mathfrak{G}^{a\nu}_{i0}&= -\frac{1}{2}(\delta^i_\lambda\mathfrak{G}^{a\nu}_{i 0}-\frac{1}{6}\mathfrak{G}^{a\nu}_{i0}R_{\lambda l~m}^{~~i}X^lX^m)(R_{~a~n}^{\lambda~0}+R_{~n~a}^{\lambda~0})X^n\\
\omega^{ij}_a\mathfrak{G}^{a\nu}_{ij}&= -\frac{1}{6}(\delta^i_\lambda\mathfrak{G}^{a\nu}_{i j}-\frac{1}{6}\mathfrak{G}^{a\nu}_{ij}R_{\lambda l~m}^{~~i}X^lX^m)(R_{~a~n}^{\lambda~j}+R_{~n~a}^{\lambda~j})X^n
\end{align}
\end{subequations}
We note in the above that the dependence of the `gravito-magnetic potential' $\frac{i}{2}\vec U$ on the Riemann tensor is linear to first order in $X^j$. This has an interesting similarity to electromagnetism. In the case of spinors in an electromagnetic field, the electromagnetic $4$-vector potential can be written as $\frac{1}{2}F^{\mu\nu}X_\nu$ when $F^{\mu\nu}$ is constant. As shown above, a similar solution holds under a gravitational field to the lowest order of spacetime dependence when the components of the Riemann tensor are constant. Indeed, the existence of a nonvanishing Riemann tensor only reveals the significance of general relativity and hence the presence of gravito-magnetic potential in spinor fields.
 
\subsection{Equatorial plane of the Kerr spacetime}
The Kerr metric, in Boyer-Lindquist coordinates, is given by \cite{Kerr}
\begin{equation}
g_{\mu\nu}=\begin{pmatrix}
\frac{\Delta-a^2\sin^2\theta}{\rho^2}&0&0&\frac{2Mar\sin^2\theta}{\rho^2}\\
0&-\frac{\rho^2}{\Delta}&0&0\\
0&0&-\rho^2&0\\
\frac{2Mar\sin^2\theta}{\rho^2}&0&0&\frac{a^2\sin^4\theta\Delta+\sin^2\theta\left(r^2+a^2\right)^2}{\rho^2}
\end{pmatrix}
\end{equation}
where $M$ is the mass of the gravitational body, $a$ the angular momentum per unit mass  and
\begin{align*}
\Delta&=r^2-2Mr+a^2\\
\rho^2&=r^2+a^2\cos^2\theta
\end{align*}
On the equatorial plane, $\theta=\pi/2$ and the first derivative of any function of the metric with respect to $\theta$ is zero. The Kerr metric then becomes 
\begin{equation}
g_{\mu\nu}=\begin{pmatrix}
1-\frac{2M}{r}&0&0&\frac{2Ma}{r}\\
0&-\frac{r^2}{\Delta}&0&0\\
0&0&-r^2&0\\
\frac{2Ma}{r}&0&0&-\left(r^2+a^2+\frac{2Ma^2}{r}\right)
\end{pmatrix}
\end{equation}
We choose the following vierbein
\begin{subequations}
\begin{align}
e^t_0&=\frac{1}{\sqrt{1-2M/r}}\\
e^r_1&=\frac{\sqrt{\Delta}}{r}\\
e^\theta_2&=\frac{1}{r}\\
e^\phi_3&= \sqrt{\frac{1-2M/r}{\Delta}}\\
e^t_3&= -\frac{2Ma}{\sqrt{\Delta(\Delta-a^2)}}
\end{align}
\end{subequations}
and the rest of the components zero. The inverse vierbein is then given by
\begin{subequations}
\begin{align}
e_t^0&=\sqrt{1-\frac{2M}{r}}\\
e_r^1&=\frac{r}{\sqrt{\Delta}}\\
e_\theta^2&=r\\
e_\phi^3&= \sqrt{\frac{\Delta}{1-2M/r}}\\
e_\phi^0&=\frac{2Ma}{\sqrt{\Delta-a^2}}
\end{align}
\end{subequations}
and the rest of the components zero.
\\

In terms of the notation introduced in the previous subsection, as $\mathfrak{F}^{KN}_{IJ}$ is antisymmetric in the indices $I$, $J$ and $K$ (once they are all lowered using the Minkowski metric) and the first derivatives with respect to $t$, $\theta$ and $\phi$ are zero, the only nonzero Christoffel symbols are $\Gamma^t_{tr}$, $\Gamma^t_{rt}$, $\Gamma^t_{r\phi}$, $\Gamma^t_{\phi r}$, $\Gamma^r_{tt}$, $\Gamma^r_{t\phi}$, $\Gamma^r_{\phi t}$, $\Gamma^r_{rr}$, $\Gamma^r_{\theta\theta}$, $\Gamma^r_{\phi\phi}$, $\Gamma^\theta_{r\theta}$, $\Gamma^\theta_{\theta r}$, $\Gamma^\phi_{tr}$, $\Gamma^\phi_{rt}$, $\Gamma^\phi_{r\phi}$ and $\Gamma^\phi_{\phi r}$. We then have
\begin{equation}
\mathrm{Re}\left(\frac{i}{2}\vec{U}\right)^j=\omega^{IJ}_\mu e^\mu_K e^\nu_N\mathfrak{F}^{KN}_{IJ}=\left\{
\begin{array}{l}
-\left(e^0_te^t_{3,r}+e^0_\phi e^\phi_{3,r}\right)e^r_1e^j_N\mathfrak{F}^{1N}_{03}\\
-\left(e^1_re^t_3\Gamma^r_{tt}+e^1_re^\phi_3\Gamma^r_{\phi t}-e^3_\phi e^r_1\Gamma^\phi_{rt}\right)e^t_0e^j_N\mathfrak{F}^{0N}_{13}\\
-\left(e^0_te^r_1\Gamma^t_{r t}+e^1_re^t_0\Gamma^r_{tt}+e^0_\phi e^r_1\Gamma^\phi_{rt}\right)e^t_3e^j_N\mathfrak{F}^{3N}_{01}\\
-\left(e^0_te^t_3\Gamma^t_{t r}+e^0_te^\phi_3\Gamma^t_{\phi r}+e^0_\phi e^t_3\Gamma^\phi_{tr}+e^3_\phi e^t_0\Gamma^\phi_{tr}+e^0_\phi e^\phi_3\Gamma^\phi_{\phi r}\right)e^r_1e^j_N\mathfrak{F}^{1N}_{03}\\
-\left(e^0_te^r_1\Gamma^t_{r \phi}+e^1_re^t_0\Gamma^r_{t\phi}+e^0_\phi e^r_1\Gamma^\phi_{r\phi}\right)e^\phi_3e^j_N\mathfrak{F}^{3N}_{01}
\end{array}
\right.
\end{equation}
where the derivatives with respect to $r$ are given by
\begin{subequations}
\begin{align}
e^t_{3,r}&=\frac{2Ma(r-M)(2\Delta-a^2)}{[\Delta(\Delta-a^2)]^\frac{3}{2}}\\
e^\phi_{3,r}&=\frac{3M\Delta-\Delta r -Ma^2}{\sqrt{\Delta^3(\Delta-a^2)}}
\end{align}
\end{subequations} 
and the Christoffel symbols are given by
\begin{subequations}
\begin{align}
\Gamma^t_{tr}=\Gamma^t_{rt}&=\frac{M}{\Delta}\left(1+\frac{a^2}{r^2}\right)\\
\Gamma^t_{r\phi}=\Gamma^t_{\phi r}&=\frac{Ma}{\Delta}\left(3+\frac{a^2}{r^2}\right)\\
\Gamma^r_{tt}&=\frac{M\Delta}{r^4}\\
\Gamma^r_{t\phi}=\Gamma^r_{\phi t}&=-\frac{Ma\Delta}{r^4}\\
\Gamma^\phi_{tr}=\Gamma^\phi_{rt}&=\frac{M}{\Delta r^2}\\
\Gamma^\phi_{r\phi}=\Gamma^\phi_{\phi r}&=\frac{1}{\Delta}\left(2M-r+\frac{Ma^2}{r^2}\right)
\end{align}
\end{subequations}
Once again, using the fact that $\mathfrak{F}^{KN}_{IJ}$ is antisymmetric in the indices $I$, $J$ and $K$, we may write $\mathrm{Re}\left(\frac{i}{2}\vec{U}\right)$ as
\begin{equation}
\mathrm{Re}\left(\frac{i}{2}\vec{U}\right)^j=\mathfrak{E}e^j_N\mathfrak{F}^{3N}_{01}
\end{equation}
where
\begin{equation}
\mathfrak{E}=\left\{
\begin{array}{l}
\left(e^0_te^t_{3,r}+e^0_\phi e^\phi_{3,r}\right)e^r_1+\left(e^1_re^t_3\Gamma^r_{tt}+e^1_re^\phi_3\Gamma^r_{\phi t}-e^3_\phi e^r_1\Gamma^\phi_{rt}\right)e^t_0\\
-\left(e^0_te^r_1\Gamma^t_{r t}+e^1_re^t_0\Gamma^r_{tt}+e^0_\phi e^r_1\Gamma^\phi_{rt}\right)e^t_3\\
+\left(e^0_te^t_3\Gamma^t_{t r}+e^0_te^\phi_3\Gamma^t_{\phi r}+e^0_\phi e^t_3\Gamma^\phi_{tr}+e^3_\phi e^t_0\Gamma^\phi_{tr}+e^0_\phi e^\phi_3\Gamma^\phi_{\phi r}\right)e^r_1\\
-\left(e^0_te^r_1\Gamma^t_{r \phi}+e^1_re^t_0\Gamma^r_{t\phi}+e^0_\phi e^r_1\Gamma^\phi_{r\phi}\right)e^\phi_3
\end{array}
\right.
\end{equation}
The individual components are given by
\begin{subequations}
\begin{align}
\mathrm{Re}\left(\frac{i}{2}\vec{U}\right)^r&=\mathfrak{E}e^r_1\mathfrak{F}^{31}_{01}\\
\mathrm{Re}\left(\frac{i}{2}\vec{U}\right)^\theta&=\mathfrak{E}e^\theta_2\mathfrak{F}^{32}_{01}\\
\mathrm{Re}\left(\frac{i}{2}\vec{U}\right)^\phi&=\mathfrak{E}e^\phi_3\mathfrak{F}^{33}_{01}
\end{align}
\end{subequations}
\section{Summary}
The attempt to understand gravity from a gauge perspective is certainly no new pursuit. Indeed, Utiyama  \cite{Utiyama} addressed the problem soon after Yang and Mills' path-breaking work on $SU(N)$ gauge groups \cite{Yang-Mills}. In Utiyama's approach, the full Poincar\'e group of Killing isometries of spacetime was utilised as a gauge group, with the Lorentz degrees of freedom being attributed to the spinor connection and translational degrees of freedom to the vierbein. In contrast, we took only the Lorentz group as the gauge group and introduced the vierbein as mathematical constructs. It was only later that they were shown to be equivalent to the usual notion of vierbein fields so that they comply with the requirement that the field Lagrangian density be the simplest nontrivial gauge-invariant diffeomorphism-invariant scalar possible, in the sense we had defined earlier. 
\\

Additionally, we have investigated the outcome of the above exercise in various limits, 
and situations such as the weak field and Schr\"odinger limits, Fermi normal coordinates and the vicinity of a Kerr body, of which most celestial objects are a good approximation. It is hoped that this shall felicitate the experimental verification of the results regarding the behaviour of spin-1/2 particles in gravitational fields that we have obtained.
\section*{Appendix}
We now present the proof of the following result
\begin{equation*}
\left[\sigma_{IJ},\sigma_{KL}\right] = -2i\delta^{PA}_{IJ}\delta^{QB}_{KL}\eta_{PQ}\sigma_{AB}
\end{equation*}
\emph{Proof:} Consider first the commutator 
\begin{align*}
\left[\gamma_I\gamma_J,\gamma_K\gamma_L\right]&= \gamma_I\gamma_J\gamma_K\gamma_L - \gamma_K\gamma_L\gamma_I\gamma_J\\
&= \gamma_I\left(2\eta_{JK}-\gamma_K\gamma_J\right)\gamma_L - \gamma_K\left(2\eta_{LI}-\gamma_I\gamma_L\right)\gamma_J\\
&= 2\eta_{JK}\gamma_I\gamma_L - 2\eta_{LI}\gamma_K\gamma_J -\gamma_I\gamma_K\gamma_J\gamma_L+\gamma_K\gamma_I\gamma_L\gamma_J\\
&= 2\eta_{JK}\gamma_I\gamma_L - 2\eta_{LI}\gamma_K\gamma_J \\
&\qquad +\frac{1}{2}\left(-\gamma_I\gamma_K\gamma_J\gamma_L-\gamma_K\gamma_I\gamma_J\gamma_L+\gamma_K\gamma_I\gamma_J\gamma_L+\gamma_K\gamma_I\gamma_L\gamma_J\right)\\
&\qquad +\frac{1}{2}\left(-\gamma_I\gamma_K\gamma_J\gamma_L-\gamma_I\gamma_K\gamma_L\gamma_J+\gamma_I\gamma_K\gamma_L\gamma_J+\gamma_K\gamma_I\gamma_L\gamma_J\right)\\
&=2\eta_{JK}\gamma_I\gamma_L - 2\eta_{LI}\gamma_K\gamma_J + \eta_{IK}\left[\gamma_L,\gamma_J\right]+\eta_{LJ}\left[\gamma_K,\gamma_I\right]
\end{align*}
Now, as the commutator bracket is linear in both its arguments
\begin{align*}
\left[\sigma_{IJ},\sigma_{KL}\right]&=\left[\frac{i}{2}\left[\gamma_I,\gamma_J\right],\frac{i}{2}\left[\gamma_K,\gamma_L\right]\right]\\
&= -\frac{1}{4}\left[\gamma_I\gamma_J-\gamma_J\gamma_I,\gamma_K\gamma_L-\gamma_L\gamma_K\right]\\
&= -\frac{1}{4}\left(\left[\gamma_I\gamma_J,\gamma_K\gamma_L\right]-\left[\gamma_J\gamma_I,\gamma_K\gamma_L\right]-\left[\gamma_I\gamma_J,\gamma_L\gamma_K\right]+\left[\gamma_J\gamma_I,\gamma_L\gamma_K\right]\right)\\
&= -\frac{1}{4}\left(\left[\gamma_I\gamma_J,\gamma_K\gamma_L\right]-\left[\gamma_J\gamma_I,\gamma_K\gamma_L\right]+\left[\gamma_L\gamma_K,\gamma_I\gamma_J\right]-\left[\gamma_L\gamma_K,\gamma_J\gamma_I\right]\right)\\
&= -\frac{1}{4}\left(2\eta_{JK}\gamma_I\gamma_L - 2\eta_{LI}\gamma_K\gamma_J + \eta_{IK}\left[\gamma_L,\gamma_J\right]+\eta_{LJ}\left[\gamma_K,\gamma_I\right]\right.\\
&\qquad -2\eta_{IK}\gamma_J\gamma_L + 2\eta_{LJ}\gamma_K\gamma_I - \eta_{JK}\left[\gamma_L,\gamma_I\right]-\eta_{LI}\left[\gamma_K,\gamma_J\right]\\
&\qquad +2\eta_{KI}\gamma_L\gamma_J - 2\eta_{JL}\gamma_I\gamma_K + \eta_{LI}\left[\gamma_J,\gamma_K\right]+\eta_{JK}\left[\gamma_I,\gamma_L\right]\\
&\qquad \left.-2\eta_{KJ}\gamma_L\gamma_I + 2\eta_{IL}\gamma_J\gamma_K - \eta_{LJ}\left[\gamma_I,\gamma_K\right]-\eta_{IK}\left[\gamma_J,\gamma_L\right]\right)\\
&= \eta_{IK}\left[\gamma_J,\gamma_L\right]-\eta_{JK}\left[\gamma_I,\gamma_L\right]+\eta_{IL}\left[\gamma_J,\gamma_K\right]+\eta_{JL}\left[\gamma_I,\gamma_K\right]\\
&= \delta^{PA}_{IJ}\delta^{QB}_{KL}\eta_{PQ}\left[\gamma_A,\gamma_B\right]\\
&= -2i\delta^{PA}_{IJ}\delta^{QB}_{KL}\eta_{PQ}\sigma_{AB}
\end{align*}
as was to be shown. 

\end{document}